\documentclass[conference, 10pt]{IEEEtran}

\usepackage{bm, latexsym, amsmath, amssymb, times}
\usepackage{cite, setspace, authblk, color, url}
\usepackage{rotating, multirow, booktabs, psfrag, epsfig, wrapfig, svg}
\usepackage[caption=false, font=footnotesize]{subfig}
\usepackage{theorem, verbatim, algorithm, algpseudocode, algorithmicx}

\IEEEoverridecommandlockouts

\begin{document}

\title{An Adaptive Framework for Generalizing Network Traffic Prediction towards Uncertain Environments}

    \author[1]{Alexander Downey}
    \author[2]{Evren Tuna}
    \author[1]{Alkan Soysal}
    \affil[1]{\normalsize Wireless@VT, Department of Electrical and Computer Engineering, Virginia Tech, Blacksburg, VA 24061, USA}
    \affil[2]{\normalsize Department of Research \& Development, ULAK Communications Inc., Ankara 06510, Turkey}

\maketitle

\begin{abstract}
We have developed a new framework using time-series analysis for dynamically assigning mobile network traffic prediction models in previously unseen wireless environments. Our framework selectively employs learned behaviors, outperforming any single model with over a 50\% improvement relative to current studies. More importantly, it surpasses traditional approaches without needing prior knowledge of a cell. While this paper focuses on network traffic prediction using our adaptive forecasting framework, this framework can also be applied to other machine learning applications in uncertain environments. 

The framework begins with unsupervised clustering of time-series data to identify unique trends and seasonal patterns. Subsequently, we apply supervised learning for traffic volume prediction within each cluster. This specialization towards specific traffic behaviors occurs without penalties from spatial and temporal variations. Finally, the framework adaptively assigns trained models to new, previously unseen cells. By analyzing real-time measurements of a cell, our framework intelligently selects the most suitable cluster for that cell at any given time, with cluster assignment dynamically adjusting to spatio-temporal fluctuations.
\end{abstract}

\section{Introduction}
The complexity of Next Generation (NextG) networks has significantly increased with the integration of key components like network slicing, private networks, and edge computing, each tailored for specific services. Employing Machine Learning (ML) has become imperative in enhancing the robustness, autonomy, and reliability of NextG cellular networks \cite{Shafin2020, Zhang2019}. An example is the Open Radio Access Network (O-RAN), which utilizes RAN Intelligent Controllers (RIC) for network management, operating in near-real-time (10-1000 ms) mode for tasks like network slicing and radio resource management and non-real-time (over 1 s) mode for functions such as energy efficiency management \cite{Polese2023}. Accurate network traffic prediction plays a crucial role in these operations.

Network traffic prediction is fundamentally a time-series forecasting problem, for which various models have been proposed \cite{Wang2017, Huang2017, Chen2018a, Zhang2018, Zhang2018b, Zhang2019c, Qiu2018, Feng2018, Gao2019, Bega2020, Wang2022b, Chen2023, Wang2019, Zhao2020, Fang2018, Tuna2022, Tuna2023, Tuna2023b}. While traditional solutions often rely on statistical methods, recent advances in machine learning, particularly deep learning, have shown significant promise. However, existing literature tends to be limited in scope, considering a small set of cells, and often struggles to scale up to the complexity of an entire cellular network.

\begin{figure*}[t]
\centerline{\includegraphics[width=.8\linewidth]{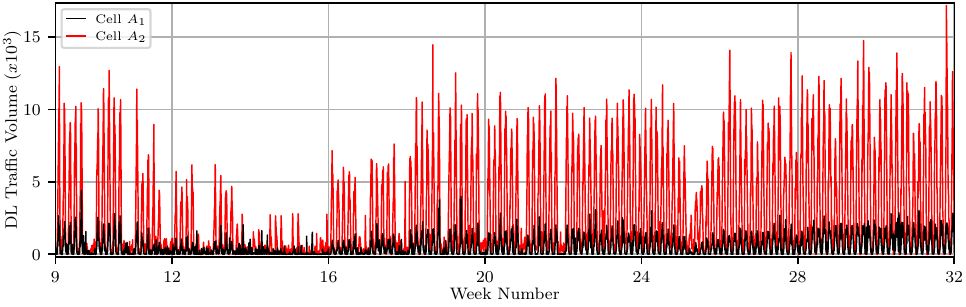}}
\caption{Traffic data taken from cells of the same base station and region. Note the large difference between the time signatures of the cells.}
\label{fig:cellData}
\vspace{-12pt}
\end{figure*}

The scalability of network traffic prediction to large networks, encompassing hundreds or even thousands of individual cell signals, poses a significant challenge. Prior work is often ill-suited for such extensive characterization. A common but naïve approach is cell-based granularity, where each cell in the network has a dedicated model \cite{Zhao2020, Fang2018, Tuna2022, Tuna2023, Tuna2023b}. While intuitive, this method necessitates training and maintaining an impractical number of models, leading to prohibitive costs in real-world applications. Additionally, the reliability of these models hinges on having sufficient data from each cell, a condition not always met in newly deployed cell locations. Due to the unique time signatures of each cell within a provider's network (see Fig.~\ref{fig:cellData}), models trained on one cell often perform poorly on others. This indicates that training with cell-based granularity fails to account for broader system interactions. Other approaches attempt to overcome this by training models on grouped data, either by base station \cite{Qiu2018, Feng2018, Gao2019, Bega2020, Wang2022b, Chen2023, Wang2019} or geographic regions  \cite{Wang2017, Huang2017, Chen2018a, Zhang2018, Zhang2018b, Zhang2019c}. While this improves predictions within those groupings, the models still struggle to generalize to cells outside their training set and adapt poorly to changing network conditions. This limitation underscores that geographic proximity alone is insufficient for comprehensive cell characterization (see Fig. \ref{fig:cellData}), highlighting the need for more versatile approaches in network traffic prediction.

The challenge of generalizing network traffic prediction among cells is further complicated by temporal dynamics. Existing models\cite{Wang2017, Huang2017, Chen2018a, Zhang2018, Zhang2018b, Zhang2019c, Qiu2018, Feng2018, Gao2019, Bega2020, Wang2022b, Chen2023, Wang2019, Zhao2020, Fang2018, Tuna2022, Tuna2023, Tuna2023b} often fail to adequately address the dynamic nature of cell tower interactions and behaviors, which are crucial for consistent and reliable predictions. These dynamics are influenced by a complex interplay of socio-technical factors, including consumer behavior, new cell deployments, and changes in local infrastructure, such as housing developments and transportation routes. These elements can drastically alter the time signatures of traffic data in a cell. A notable example is observed in our dataset during the COVID-19 pandemic (dip in traffic volumes in Fig.~\ref{fig:cellData} occurred during COVID-19 lockdowns), where models trained on pre-pandemic data underperformed when forecasting traffic during the pandemic, and vice versa. This illustrates how shifts in the underlying infrastructure and user demand lead to significant changes in network traffic, highlighting the necessity for predictive models that can adapt over time to maintain accuracy and effectiveness.

We propose a three-step \textit{cluster-train-adapt} framework, designed to adaptively address both temporal and spatial changes during operation, thereby overcoming the generalization issues prevalent in prior work. This framework employs an unsupervised clustering technique to separate the extraction of time-signature features from geographical and temporal cell interactions. We then integrate this with supervised prediction models, one for each cluster. This approach results in more accurate network traffic volume forecasting since each cluster focuses on a specific traffic behavior. Our multi-model architecture, capable of utilizing several prediction models simultaneously, offers a diverse ensemble that caters to various network traffic types. This approach not only improves coverage and predictive accuracy but also excels in scenarios where models have no prior knowledge of specific cells. Unlike many existing works that rely on increasing model complexity for marginal gains, our strategy leverages domain knowledge to achieve over a 50\% improvement in prediction accuracy without added model complexity. 

While the primary focus of this paper is on network traffic prediction for previously unseen cells, the underlying principles of our adaptive framework extend well beyond this specific application. They hold significant potential for generalizing any machine learning model to uncertain environments, particularly in scenarios where the test data originates from a different probability distribution than the training dataset. This aspect of our work underscores its relevance in a wide range of applications where machine learning models must adapt to evolving or previously unknown conditions.

\section{Cluster-Train-Adapt Framework}

\begin{figure}[t]
    \centering 
    \includegraphics[width =.85\columnwidth]{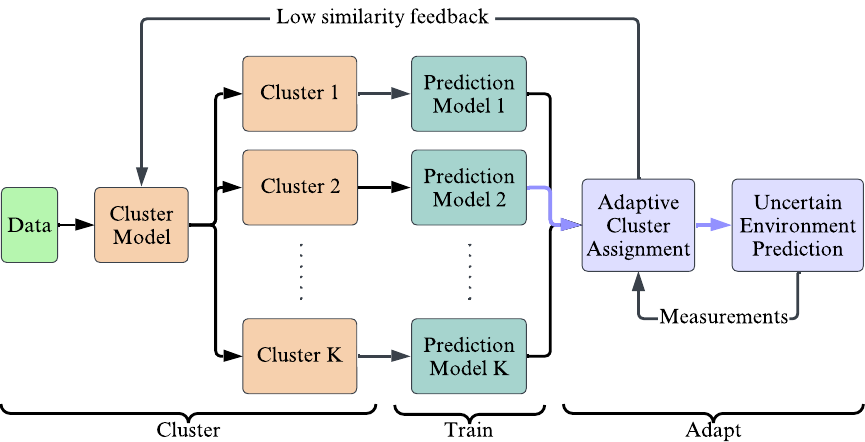} 
    \caption{Our adaptive framework includes cluster, train, and adapt steps. In an uncertain environment, adaptive cluster assignment determines the best prediction model using live measurements from the environment.}
    \label{fig:framework}
    \vspace{-12pt}
\end{figure}

In this section, we introduce our innovative adaptive framework, designed to capture network-specific behaviors using an ensemble of unsupervised and supervised learning techniques. Central to our approach is the ability to adapt to the unique characteristics of different environments. The framework initiates with unsupervised clustering of time-series data, aiming to identify distinct trends and seasonal patterns inherent in the training data. This is followed by applying supervised learning techniques, tailored to predict the output feature within each identified cluster. The final step involves adaptively assigning these trained models to new, previously unseen environments, ensuring optimal performance across a spectrum of uncertainties. Fig.~\ref{fig:framework} provides a visual representation of these three critical steps, illustrating how each component contributes to the overall effectiveness of our framework. 

In the following section, we demonstrate the application of this framework to the specific challenge of network traffic prediction, showcasing its versatility and efficacy in handling spatio-temporal dynamics in a large mobile network.

\subsection{Clustering}
The clustering step of our framework starts with data preprocessing, where we first determine the smallest size seasonality, denoted by $n$, of the time-series data. As an example, for network traffic prediction, data has a 24-hour seasonality, $n=24$, as can be observed from Fig.~\ref{fig:cellData}. We then partition the data into smaller time-series segments of length $n$, ensuring that each segment encapsulates a complete seasonal cycle. In scenarios involving multiple time series, we consolidate these segments to form a unified dataset. In the case of network traffic prediction, this corresponds to consolidating segments from different cells. 

Upon completing the preprocessing, we proceed to cluster the consolidated dataset into $K$ distinct groups using a chosen clustering model and an appropriate similarity measure. This clustering process effectively groups time series segments with analogous time signatures, allowing us to isolate and categorize distinct traffic patterns. Following this, the framework transitions into the next critical stage: training a dedicated prediction model for each of the $K$ clusters.

\subsection{Training Prediction Models}
In the training step of our framework, we train distinct models for each cluster separately using the data specific to cluster $k$. The prediction architecture uses $n$ past values to predict $m$ steps ahead. This training approach is advantageous as it significantly narrows the variation within each cluster, thereby enabling the development of more accurate models. These models are not just more precise but are also better suited for the unique traffic characteristics inherent within each dataset cluster. 

\subsection{Adapting to Uncertain Environments}
In the adapting step of our framework, we test our predictive models on previously unseen time series data. Central to this process is the evaluation of live measurements using the same similarity score employed in the clustering step. This enables the framework to intelligently assign the most appropriate cluster and consequently, the best predictive model to each test series. Given that each model is trained on data with similar characteristics, it is inherently better equipped for accurate predictions.

The framework continuously reevaluates measurement feedback every $n$ steps. This dynamic reevaluation process allows for the reassignment of a better-suited predictive model in response to any changes in the temporal characteristics of the test series. When applied to an entire sequence of network traffic data, this approach ensures that the most effective model is assigned for each distinct spatio-temporal behavior, significantly enhancing the framework's ability to generalize effectively compared to traditional methods.

Furthermore, our adaptive framework incorporates a larger time-scale feedback loop to adeptly manage any uncertain environments not represented in the initial training dataset. During the adaptive cluster assignment, if the similarity score falls below a certain threshold, indicating a potential out-of-distribution scenario, the framework initiates the collection of data from the test series. This data is then added to the existing training set. Once a substantial volume of out-of-distribution data is amassed, we reapply the clustering algorithm to form a new cluster, $K+1$. This innovative approach not only adapts to the current environment, but also evolves the framework's capacity to handle new, unforeseen scenarios, continuously enhancing its predictive prowess over time.

\section{Mobile Network Traffic Prediction}
In this section, we demonstrate the application of our proposed adaptive framework to the domain of mobile network traffic prediction. Our primary objective is to validate the effectiveness of our framework, and for this purpose, we employ established clustering and traffic prediction models from existing literature. Specifically, we utilize a network traffic dataset from \cite{Tuna2023b}, a time-series K-means clustering algorithm with a dynamic time warping (DTW)-based similarity score \cite{Tavenard2020}, and an LSTM-based deep learning model for prediction \cite{Tuna2023b}. 

As a benchmark for our study, we consider the single-cell analysis results from \cite{Tuna2023b} as our baseline. In their single-cell analysis, a multivariate LSTM model is trained on the first 40 weeks of data and tested on the final 4 weeks out of a total of 52 weeks from a selected cell. Distinct from this approach, our framework is designed to not see any data from this specific cell during training. Instead, it learns from the data of other cells in the dataset and is then tested on the same selected cell.   

\subsection{Dataset}

In our study, we utilize the dataset described in \cite{Tuna2023b}, which compromises 138 unique carrier cells, each containing 52 weeks worth of hourly aggregated data. Each data point includes 20 distinct radio access network (RAN) features, such as downlink traffic volume, average number of active users in downlink, and downlink PRB utilization. Similar to \cite{Tuna2023b}, we focus on the downlink traffic volume as our output feature. 

Given the dataset's inherent 24-hour seasonality, our model uses the data from the first 24 hours as the input window for prediction. We aim to forecast one step ahead. In our training process, we deliberately exclude the cell labeled as $A_2$ in \cite{Tuna2023b}.

\subsection{Dynamic Time Warping}

\begin{figure}[t]
    \centering 
    \includegraphics[width = \columnwidth]{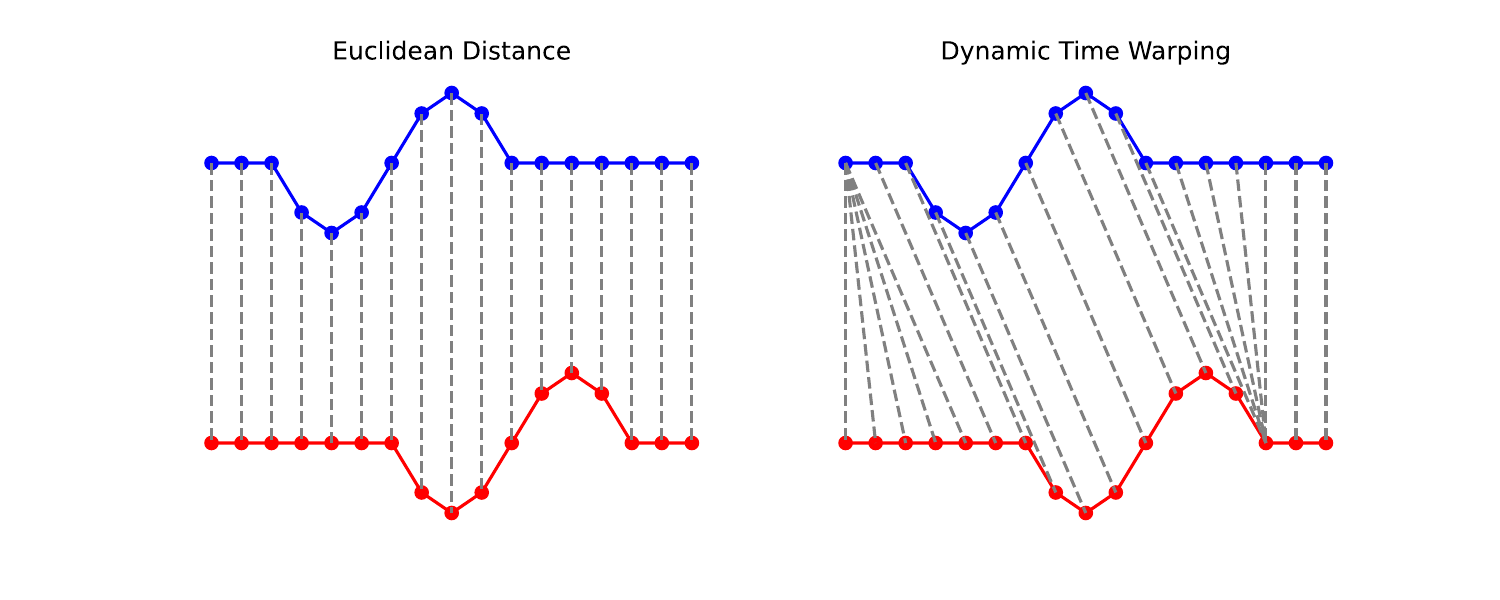} 
    \vspace{-30pt}
    \caption{The difference in alignment between Euclidean distance and DTW.}
    \label{fig:dtw}
    \vspace{-12pt}
\end{figure}
Several studies pertain to the application of unsupervised learning towards clustering time series data. Popular approaches adapt the methodologies from K-means clustering where the \textit{distance metric} and \textit{centroid representations} are designed to accommodate time series data instead of a singular value. The authors in \cite{Tavenard2020} provide such an implementation that uses DTW, providing better invariance to changes in frequency and magnitude when characterizing over Euclidean distance. Given two time series $x$ and $x'$, DTW selects a mapping $\pi$ from the set of all admissible mappings $A(x,x^\prime)$ to minimize the Euclidean distances between the two aligned time series, with some sensitivity $q$. Fig.~\ref{fig:dtw} provides a conceptual example of this alignment. The DTW equation is given as 
\begin{align}
    {DTW}_q{(x,x^\prime)}=\min_{\pi\in A(x,x^\prime)}{\left(\sum_{(i,j)\in\pi}{d(x_i,{x^\prime}_j)}^q\right)^\frac{1}{q}}
    \label{eq:dtw}
\end{align}

Similar to \cite{Tavenard2020}, we use DTW as the \textit{distance metric} and utilize DTW barycenter averaging (DBA) to form \textit{centroid representations}. Centroid formulation is given as
\begin{equation}
    {DBA}{(C)}=\arg\!\min_{c^\prime\in X}{\left(\sum_{c\in C}{DTW(c,c^\prime)}\right)} ; C \subseteq X
\end{equation}
where a centroid $c^\prime$ is selected to minimize the sum of the DTW distances for a given cluster $C$ where $X$ is a superset of all time-series sequences. In this manner, we can perform K-means clustering on time-series data.

\subsection{Network Traffic Prediction Models}
The findings of \cite{Tuna2023b} devote a considerable portion of work towards the analysis, selection, and engineering of this dataset to improve its predictability in a deep learning context, finding several feature configurations to evaluate their approach. While it is likely several other configurations may prove themselves comparable, or better, candidates towards such an approach, we will use those found in \cite{Tuna2023b} because they provide a baseline for direct comparison with our results. 

\begin{figure}[t]
    \centering 
    \includegraphics[width = .8\columnwidth]{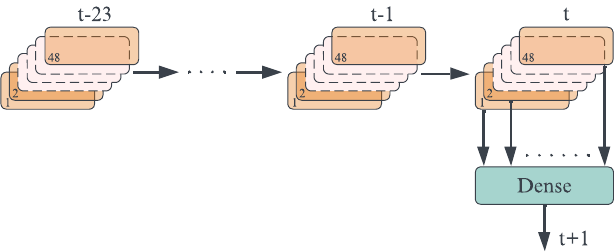} 
    \caption{LSTM model has a window size of 24 with 48 hidden layers, followed by a dense layer.}
    \label{fig:lstm}
    \vspace{-12pt}
\end{figure}

We use a series of LSTM models that are proposed in \cite{Tuna2023b} and illustrated in Fig.~\ref{fig:lstm}. The input is the past 24 hours of a set of features, predicting the next instance of downlink traffic volume. Five LSTM-based model configurations are proposed in \cite{Tuna2023b}, denoted as \textit{LSTM-uni}, \textit{LSTM-RAN}, \textit{LSTM-peak}, \textit{LSTM-handover}, and \textit{LSTM-all}. The first configuration, \textit{LSTM-uni}, considers the univariate case where the output, downlink traffic volume, is the only feature considered for the model. The second configuration, \textit{LSTM-RAN}, considers a multivariate case with five features, selecting those RAN features in the dataset that have a Pearson correlation coefficient greater than 0.9 with the downlink traffic volume. The third configuration, \textit{LSTM-peak}, considers two engineered features in addition to the downlink traffic volume. The first feature draws a distinction between peak and non-peak hours for the downlink traffic volume. The second feature establishes a binary distribution between weekdays and weekends. The fourth configuration, \textit{LSTM-handover}, considers another aspect of feature engineering by exploiting the handover relationship between different cells. This configuration adds two additional features, one for incoming handovers and the other for outgoing handovers. The final configuration, \textit{LSTM-all}, draws from all prior features. 

\subsection{Adapting to an Unseen Cell}
\begin{figure}[t]
    \centering 
    \includegraphics[width = .8\linewidth]{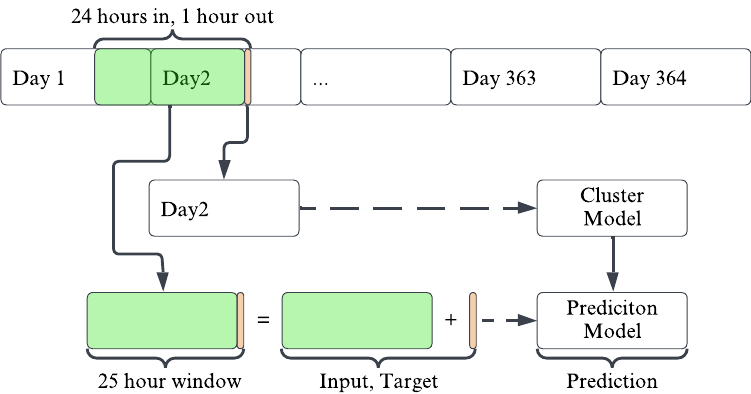} 
    \caption{Application of adaptive cluster assignment in network traffic prediction.}
    \label{fig:adapting}
    \vspace{-12pt}
\end{figure}
In this section, we present our adaptive cluster assignment procedure. In the first row of Fig.~\ref{fig:adapting}, the sliding window includes the past 24 hours (labeled with green color) and the subsequent hour, i.e., the prediction target (labeled in orange color). In the second row of Fig.~\ref{fig:adapting}, when the prediction target falls on Day 2, our framework evaluates which cluster Day 2 belongs to using the DTW score. Let us assume that Day 2 is most similar to the centroid of Cluster $K$, then the prediction model trained with Cluster $K$ data is assigned to the test cell. In the third row of Fig.~\ref{fig:adapting}, the 25-hour window is divided into a 24-hour input and a target. The prediction model of Cluster $K$, together with 24-hour past data is used to predict the subsequent hour. When the sliding window moves to Day 3, the framework will reevaluate the cluster assignment. If Day 3 belongs to another cluster, the framework assigns another prediction model to the test cell. 
\begin{figure}[t]
\centering
\subfloat[Trained with MAE loss using one cluster.]{\includegraphics[width=.85\columnwidth]{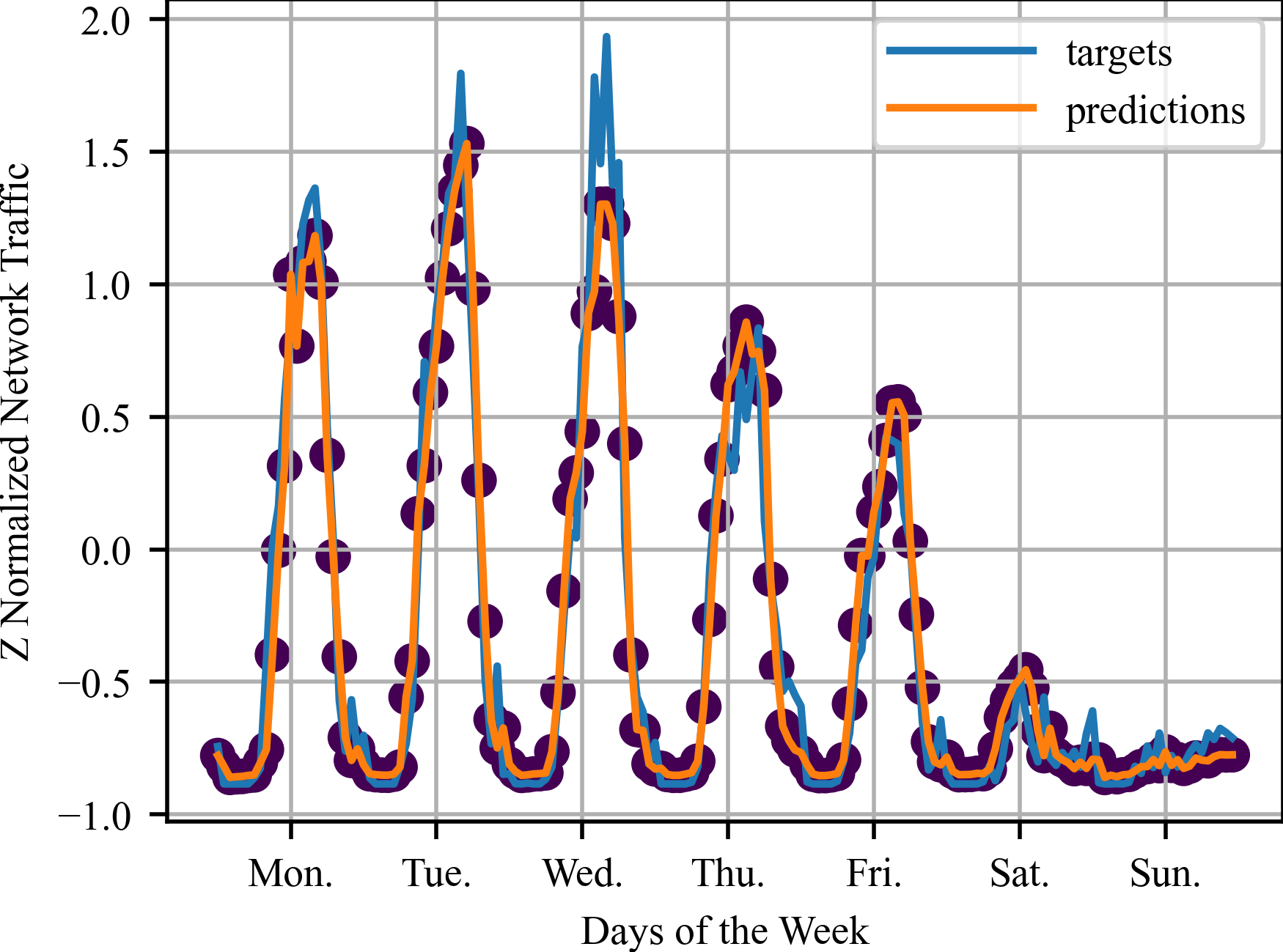}}\\
\subfloat[Trained with MAE loss using eight clusters.]{\includegraphics[width=.85\columnwidth]{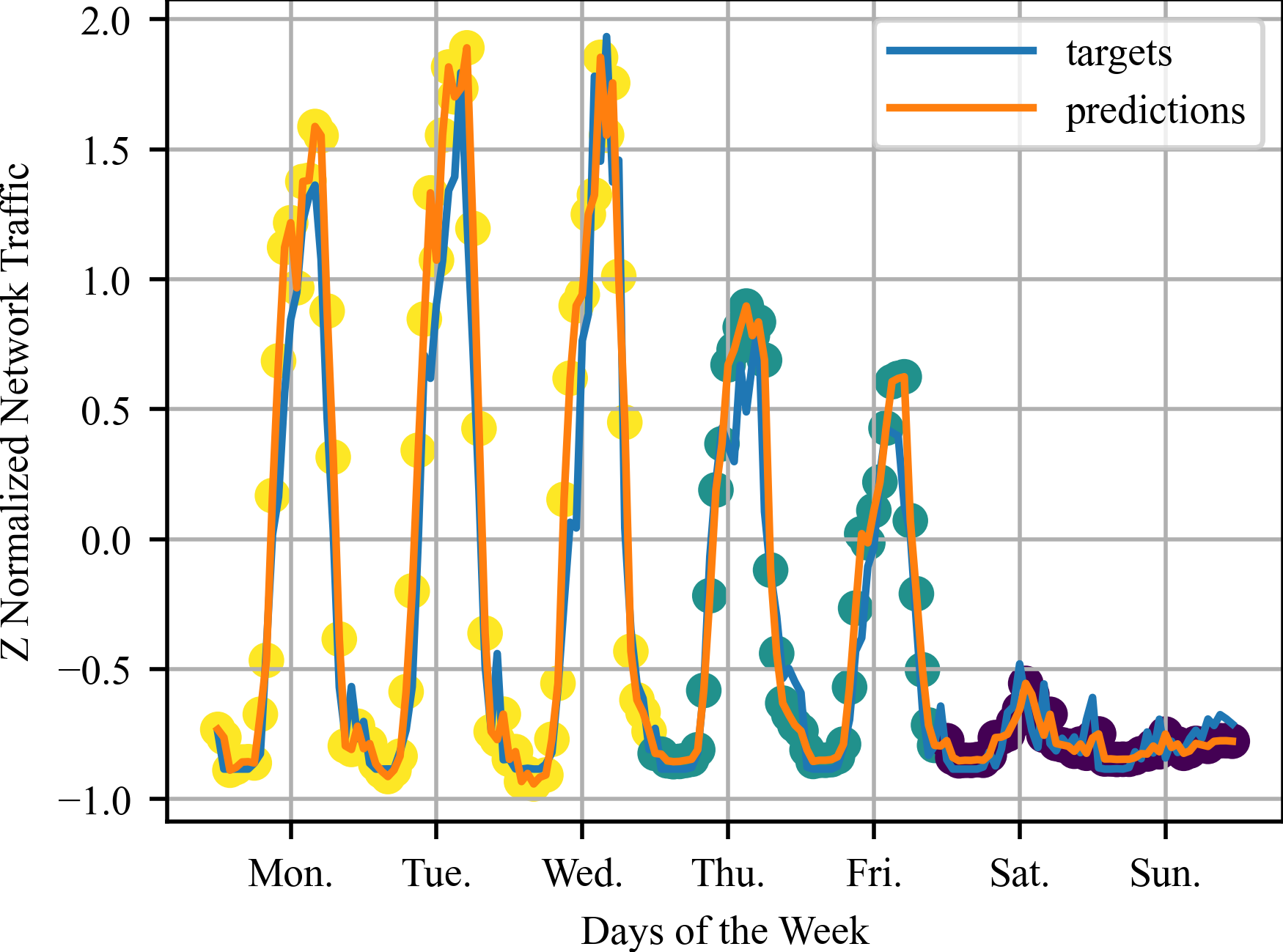}}
\caption{Plot of predictions in blue compared to ground truth in orange. The color of the dots represents the cluster of each data point.}
\label{fig:time_plots}
\vspace{-12pt}
\end{figure}

In this paper, we do not implement the larger time-scale feedback loop. Instead, we run our framework with multiple $K$ values. We plot our test cell predictions using $K=1$ in Fig.~\ref{fig:time_plots}(a) and $K=8$ in Fig.~\ref{fig:time_plots}(b). We observe in Fig.~\ref{fig:time_plots}(b) that the assigned cluster changes on Thursday and again on Saturday to accommodate the change in daily fluctuations.

\section{Experiments and Results}
\label{sec:ER}

Clustering models were trained for 100 iterations on the time series K-means implementation provided by \cite{Petitjean2011, Tavenard2020} using the DTW metric. Several K values were used to group the data into different numbers of clusters. Prediction models were trained for 90 epochs with MAE loss using the SGD optimizer with a momentum of 0.9. Training used an initial learning rate of 0.1, which was reduced on validation loss plateau (with a patience of 10) by a factor of 10. Training also used early stopping with a patience of 40. It is important to note that we train a prediction model for each cluster of data from the clustering model (e.g. if we train with 8 clusters, we train 8 prediction models).

Evaluating this approach with any values of K greater than 1 will involve several predictive models. As such, the final loss is a weighted average of each model proportional to the amount of data in each corresponding cluster.

We trained models with clusters of $K= 1, 2, 4, 8,$ and $16$ with MAE loss under 5 different feature configurations: \textit{LSTM-uni}, \textit{LSTM-RAN}, \textit{LSTM-peak}, \textit{LSTM-handover}, and \textit{LSTM-all}. We evaluated models with the entirety of Cell $A_2$ and a subset of Cell $A_2$, the four test weeks used in \cite{Tuna2023b}. 

\begin{figure}[t]
\centerline{\includegraphics[width=\columnwidth]{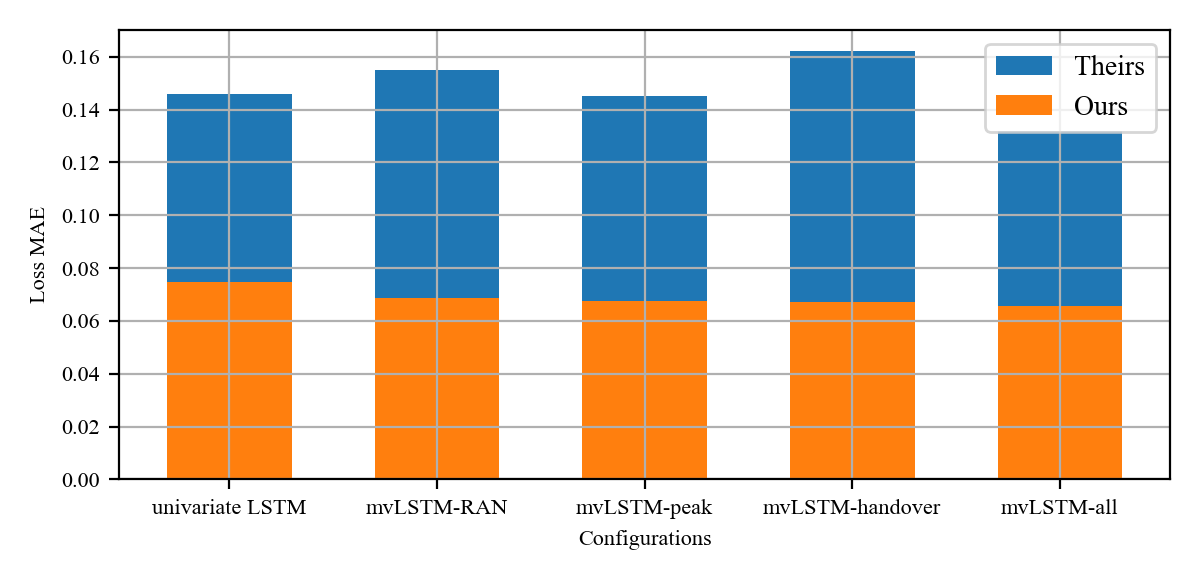}}
\caption{Chart of MAE loss values as a function of feature configuration for a cluster size of 8 compared to the results of \cite{Tuna2023b}. Taken from Table~\ref{tab:MAEsubres}.}
\label{fig:wMAEbar}
\vspace{-12pt}
\end{figure}

In order to compare the proposed framework to \cite{Tuna2023b}, we first present our results on the subset of the test cell $A_2$ in Table~\ref{tab:MAEsubres}. It is important to note that \textit{LSTM-all} configuration in \cite{Tuna2023b} is the state-of-the-art result and performs significantly better than other relevant literature. \textit{LSTM-all} configuration in \cite{Tuna2023b} has an MAE loss of 0.132, while our framework with $K=8$ has an MAE loss of 0.065. Our proposed adaptive framework approach results in a 50\% less MAE loss than the best result in \cite{Tuna2023b}. When using a sufficiently large number of clusters (e.g. 8), we find that our approach outperforms all feature configurations of \cite{Tuna2023b} as shown in Fig.~\ref{fig:wMAEbar}.

\begin{table}[t]
\caption{MAE Loss Results on Subset of Cell $A_2$}
\footnotesize
\begin{center}
\begin{tabular}{l|c|c|c|c|c|c}
\toprule
\textbf{\textit{Configurations}}& \textbf{\textit{${K}_1$}}& \textbf{\textit{${K}_2$}}& \textbf{\textit{${K}_4$}}& \textbf{\textit{${K}_8$}}& \textbf{\textit{${K}_{16}$}} & \textit{\cite{Tuna2023b}}\\
\midrule
\textit{LSTM-uni}      &.126 &.149 &.190 &.075 &\textbf{.067} &.146 \\
\textit{LSTM-RAN}      &.125 &.150 &.169 &\textbf{.069} &\textbf{.069} &.155 \\
\textit{LSTM-peak}     &.151 &.140 &.181 &.067 &\textbf{.065} &.145 \\
\textit{LSTM-handover} &.125 &.141 &.206 &\textbf{.067} &.069 &.162 \\
\textit{LSTM-all}      &.127 &.158 &.201 &\textbf{.065} &.066 &.132 \\
\bottomrule
\end{tabular}
\label{tab:MAEsubres}
\end{center}
\vspace{-12pt}
\end{table}

More importantly, our framework achieves this performance without seeing any data from the test cell $A_2$. To explore this further, we compare the performance of the approach in \cite{Tuna2023b} when the prediction model is trained using the data of a different cell and tested on cell $A_2$. Table~\ref{tab:OtherMAEResS} shows that performance improvement of using our framework can be as high as 72\%, which is achieved when compared to the prediction model trained on $B_3$ and tested on $A_2$.  

\begin{table}[t]
\caption{MAE Loss Results when trained on other cells.}
\footnotesize
\begin{center}
\begin{tabular}{c|c}
\toprule
\textbf{\textit{Trained On}}& \textbf{\textit{Results}} \\
\midrule
\textit{${A}_2$} &.146 \\
\textit{${B}_3$} &.293 \\
\textit{${C}_5$} &.230 \\
\textit{${D}_1$} &.147 \\
\textit{${E}_2$} &.149 \\
\textit{Ours} &\textbf{.067} \\
\bottomrule
\end{tabular}
\label{tab:OtherMAEResS}
\end{center}
\vspace{-12pt}
\end{table}

We attribute the larger amount of data used in our training and validation to the overall improvement in performance. It should be emphasized that $A_2$ is absent in our approach's training and validation data unlike in \cite{Tuna2023b}. The models trained by \cite{Tuna2023b} lack adequate exposure to all network behaviors in \cite{Tuna2023b}'s test set, providing poorer results. Even though we use different cells in our training, the network traffic behaviors present in the other cells span the ones in Cell $A_2$, enabling accurate model predictions. With this in mind, we note an interesting trend. While large numbers of clusters tend to improve results, too few degrade performance. We believe that the increased data in our training set over-saturates models with training examples from a variety of traffic behaviors. While this increases the generality of our approach, the lack of conditionality prevents models from learning behavior-specific features. In addition, using a small number of clusters (e.g. four clusters) separates data too generally. Models trained on these clusters lack the generality of the entire dataset while still suffering from the over-saturation of the feature space, and we see that this affects all feature configurations. It is only when we use clusters of sufficient number that we see improvements. A larger number of clusters separates data and de-saturates the feature space without degrading the generality of the approach. Each cluster contains the information necessary to characterize its portion of traffic behavior without being reliant on a larger set of behaviors to achieve good performance. In this manner, each model can specialize without over-fitting, improving results.
 
While are not directly comparable to \cite{Tuna2023b}, we saw a similar trend when the loss is calculated on the entire cell $A_2$. Table~\ref{tab:MAEres} shows that cluster sizes of $8$ or $16$ perform better than $1$, $2$, and $4$. We observe that multivariate features proposed in \cite{Tuna2023b} are still useful where \textit{LSTM-peak} configuration provides 20\% improvement compared to the univariate case of the framework.

\begin{table}[t]
\caption{MAE Loss Results on Entire Cell ${A}_2$}
\footnotesize
\begin{center}
\begin{tabular}{l|c|c|c|c|c}
\toprule
\textbf{\textit{Configurations}}& \textbf{\textit{${K}_1$}}& \textbf{\textit{${K}_2$}}& \textbf{\textit{${K}_4$}}& \textbf{\textit{${K}_8$}}& \textbf{\textit{${K}_{16}$}} \\
\midrule
\textit{LSTM-uni}      &.152 &.139 &.183 &.131 &\textbf{.127} \\
\textit{LSTM-RAN}      &.152 &.135 &.172 &\textbf{.113} &.125 \\
\textit{LSTM-peak}     &.143 &.125 &.161 &\textbf{.105} &.119 \\
\textit{LSTM-handover} &.149 &.135 &.195 &\textbf{.126} &.131 \\
\textit{LSTM-all}      &.148 &.128 &.189 &\textbf{.111} &.113 \\
\bottomrule
\end{tabular}
\label{tab:MAEres}
\end{center}
\end{table}

\begin{figure}[t]
\centerline{\includegraphics[width=.9\columnwidth]{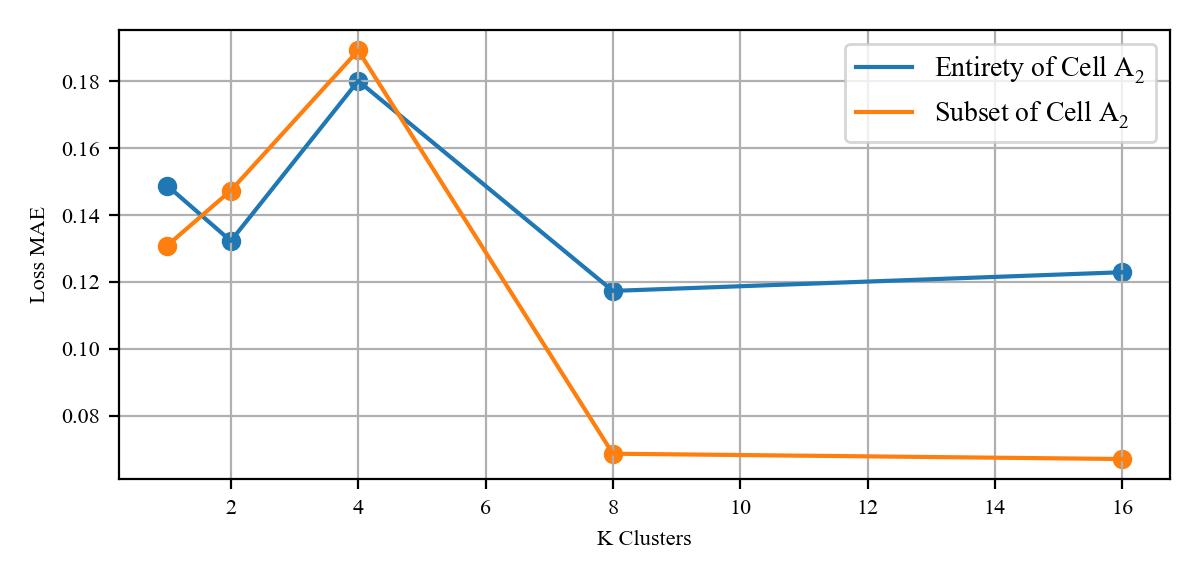}}
\caption{Plot of MAE loss values as a function of cluster size. Each value is an average of models trained on all feature configurations.}
\label{fig:MAEplot}
\vspace{-12pt}
\end{figure}

We see a difference in performance between the test cases with the entirety of \textit{$A_2$} and a subset of \textit{$A_2$} in Fig.~\ref{fig:MAEplot}. Entire \textit{$A_2$} contains a larger number of traffic behaviors, providing a more likely depiction of model results. This further reinforces our claim that prior work \cite{Tuna2023b} is not as generalizable as a subset of \textit{$A_2$} only evaluates a small sample of network behaviors. 

\section{Conclusion}
In this paper, we presented an innovative adaptive framework, termed ``cluster-train-adapt'', specifically designed for time-series data with a clear seasonality. Our framework began with an unsupervised clustering step, identifying unique trends and seasonal patterns within time-series data. This was followed by the training of distinct predictive models for each cluster. The final step involved adaptively assigning these models to previously uncertain environments, demonstrating our framework's ability to generalize and adapt to new data.

Applying our framework to mobile network traffic prediction, we successfully showcased its superiority over traditional prediction models. Notably, the 'cluster-train-adapt' approach significantly outperformed the models presented in \cite{Tuna2023b}, especially in predicting traffic for cells not encountered during training. This validates our hypothesis that a model trained on clustered data can achieve higher accuracy due to reduced intra-cluster variability.

The broader implications of our research extend well beyond cellular network traffic prediction. The principles underlying our adaptive framework can be applied to generalize various machine learning models to uncertain environments, particularly where test data differ significantly from training data. This flexibility and adaptability make our framework a valuable tool in numerous time-series forecasting applications.

\section*{Acknowledgment}
The authors acknowledge Advanced Research Computing at Virginia Tech for providing computational resources and technical support that have contributed to the results reported within this paper. URL: https://arc.vt.edu/

\end{document}